\documentclass{optica-article}
\journal{opticajournal} 
\articletype{Research Article}

\usepackage{lineno}

\usepackage{subcaption}
\usepackage{siunitx}

\makeatletter
\newcommand{\phantomlabel}[2]{
	\protected@write\@auxout{}{
		\string\newlabel{#2}{
			{\@currentlabel#1}{\thepage}
			{\@currentlabel#1}{#2}{}
		}
	}
	\hypertarget{#2}{}
}
\graphicspath{{figures/}}

\begin{document}

\title{Resonant EO combs: Beyond the standard phase noise model of frequency combs}

\author{H. R. Heebøll,\authormark{1,2,4,*} P. Sekhar,\authormark{2,3,4} J. Riebesehl,\authormark{1} \\
A. Razumov,\authormark{1} M. Heyrich,\authormark{2,3} M. Galili,\authormark{1} F. Da Ros,\authormark{1} \\
S. Diddams,\authormark{2,3} and D. Zibar\authormark{1}}

\address{\authormark{1} Department of Electrical and Photonics Engineering, Technical University of Denmark, DK-2800, Kgs. Lyngby, Denmark}
\address{\authormark{2} Electrical, Computer and Energy Engineering, University of Colorado, Boulder, CO 80309, USA}
\address{\authormark{3} Department of Physics, University of Colorado, Boulder, CO 80309, USA}
\address{\authormark{4} These authors contributed equally to this work.}
\email{\authormark{*}hohel@dtu.dk} 


\begin{abstract*} 
A resonant electro-optic (EO) frequency comb is generated through electro-optic modulation of laser light within an optical resonator. Compared to cavity-less EO combs generated in a single pass through a modulator, resonant EO combs can produce broader spectra with lower radio frequency (RF) power and offer a measure of noise filtering beyond the cavity's linewidth. Understanding, measuring, and suppressing the sources of phase noise in resonant EO combs is crucial for their applications in metrology, astrophotonics, optical clock generation, and fiber-optic communication. According to the standard phase noise model of frequency combs, only two variables—the common mode offset and repetition rate phase noise—are needed to fully describe the phase noise of comb lines. However, in this work we demonstrate analytically, numerically, and experimentally that this standard model breaks down for resonant EO combs at short timescales (high frequencies) and under certain comb parameters. Specifically, a third phase noise component emerges. Consequently, resonant EO combs feature qualitatively different phase noise from their cavity-less counterparts and may not exhibit the anticipated noise filtering. A more complete description of the deviations from the standard phase noise model is critical to accurately predict the performance of frequency combs. The description presented here provides foundational insights for improved designs tailored to applications such as supercontinuum generation and optical communication. 
\end{abstract*}

\section{Introduction}
Resonant electro-optic (EO) frequency combs avoid the multiple cascaded phase modulators
needed to achieve a broad spectrum in conventional EO combs\cite{kourogi1993wide}. Instead, they recirculate the light inside a cavity or resonator, allowing the same modulator to act multiple times before the light exits the cavity. As a result, resonant EO combs require around ten times less RF power than their cascaded counterparts \cite{sekhar202320}. Resonant EO combs are excellent sources for generating short pulses used in supercontinuum generation \cite{sekhar202320} and have applications in metrology \cite{diddams2020optical} and communications \cite{chen2022phase}.

The two key challenges in making resonant EO combs practical are implementing them on-chip and reducing their phase noise \cite{rueda2019resonant}. Recent work has explored various materials and integrated resonant EO combs on-chip \cite{yu2019chip,zhang2019broadband,lufungula2024integrated,zhang2024integrated}. However, our understanding of the phase noise in resonant EO combs remains incomplete. Earlier studies suggest that resonant EO combs follow the standard phase noise model, where only two phase noise components are present: one associated with the common offset frequency and another associated with the repetition rate frequency \cite{kim2017cavity,buscaino2020design,chen2022phase}. However, here we will confirm the recent numerical results\cite{heeboell2024resonant} predicting that resonant EO combs exhibit phase noise that is beyond this standard model.\vspace{0.2cm} 

Understanding deviations from the standard phase noise model of frequency combs is crucial, as many experimental phase-stabilization mechanisms\cite{papp2014microresonator,ideguchi2014adaptive} and digital phase-correction algorithms\cite{burghoff2019generalized,lundberg2020phase} rely on this model to suppress phase noise. 

Subspace tracking is a recently developed method that characterizes the phase correlations between comb lines. Subspace tracking can test for deviations to the standard phase noise model by decomposing the phase noise of all lines into only a few significant components. This decomposition is achieved using a multi-heterodyne digital coherent detection scheme, which measures the phase noise of all comb lines simultaneously to calculate their correlation matrix. So far, subspace tracking has confirmed that various combs follow the standard model, including the standard EO comb and a quantum-dot mode-locked laser \cite{razumov2023subspace}, as well as a Cr:ZnS-based mode-locked laser\cite{razumov2024Cr:ZnS}.\vspace{0.2cm} 

In this paper, we elaborate on recent numerical results predicting that resonant EO combs will deviate from the standard phase noise model by exhibiting a 3rd significant phase noise component \cite{heeboell2024resonant}. We propose an analytical description of this effect and confirm the results experimentally. To the best of our knowledge, this is the first phase noise component beyond the standard model that has been described analytically, numerically, and experimentally.\vspace{0.2cm}

Resonant EO comb are expected to exhibit noise filtering due to their cavity or resonator \cite{kim2017cavity,rueda2019resonant,sekhar2021fiber,bienfang2001phase}. However, in practice, a second dedicated filter cavity is often required, particularly when these combs are used for broadband supercontinuum generation \cite{sekhar2022noise,sekhar202320}. This suggests that the inherent filtering effect from the comb's cavity is smaller than anticipated. We will demonstrate that the reduced filtering is related to the 3rd phase noise component.\vspace{0.2cm} 

This paper is structured as follows. In section 2 we will present the standard phase noise model of frequency combs, comment on how resonant EO combs will deviate from this model, and introduce the main concepts behind subspace tracking. In section 3, we will numerically investigate the noise properties of resonant EO combs and subspace tracking will confirm the 3rd noise component originally predicted in \cite{heeboell2024resonant}. In Section 4, we will investigate the resonant EO comb experimentally, looking to confirm the 3rd noise component found in simulations and analytical work. We investigate how the component changes with RF power/modulation index. In Section 4, we will discuss what lessons can be learned for the design of resonant EO combs and discuss the physical intuition behind the 3rd component. \vspace{0.2cm} 
 
\section{Theory} \label{sec:theory}
\subsection{The standard model}
A frequency comb is defined as a series of spectral lines that fall on a frequency grid defined by just two variables, such that the frequency of the $ m^{\rm th} $ comb line is given by:
\begin{gather}
	f^{lines}_m = f_{0} + mf_{rep},
\end{gather}
where $f_0$ is the common offset frequency of all lines and $f_{rep}$ is the repetition rate defining the line spacing. The standard phase noise model of frequency combs describes the phase of the $ m^{\rm th} $ comb line using the two corresponding phase noise components:
\begin{gather}
	\phi^{lines}_{m}(t) = \phi_{0}(t) + m\phi_{rep}(t),
\end{gather}
where $ \phi_{0}(t) $ describes the phase noise that is common to all comb lines (corresponding to $ f_0 $), and $ \phi_{rep}(t) $ is the repetition rate phase noise (corresponding to $ f_{rep} $) whose contribution increases linearly with comb line number $ m $. This standard phase noise model has been widely studied and experimentally confirmed in various combs, including EO combs \cite{ishizawa2013phase,lundberg2018phase,parriaux2020electro}, mode-locked-lasers \cite{telle2002kerr,stenger2002ultraprecise,benkler2005circumvention,hutter2023femtosecond}, and microcombs \cite{lei2022optical}, where $ \phi_0 $ and $ \phi_{rep} $ have various physical origins depending on the exact comb design. Likewise, the combs feature different natural choices of which line is labelled $ m{=}0 $, depending on where the effect of their repetition rate phase noise vanishes. We will refer to this general form of phase noise as the standard phase noise model of frequency combs, or simply the standard model. 

Even though this standard model is very similar to the elastic tape model\cite{telle2002kerr,benkler2005circumvention}, it differs by not necessarily defining $ f_0 $ as the central envelope offset frequency (often defined as the smallest positive frequency on the grid) and does not explicitly include the free parameter of a fixed point at which the repetition rate phase noise vanished.\vspace{0.2cm} 

An example of a comb that follows the standard model is---what we will refer to as---the standard EO comb. It consists of a CW laser with frequency $ f_{CW} $ feeding into a phase modulator driven by an RF generator at a frequency of $ f_{RF} $. This produces a series of equally spaced comb lines, such that the $m^{\rm th}$ line has the frequency $ f^{lines}_m = f_{CW} + mf_{RF} $. Accordingly, the standard model of EO combs features a common contribution from the CW laser's phase noise $ \phi_{CW}(t) $ and a second contribution from the RF generator's phase noise $ \phi_{RF}(t) $ that increases linearly with $ m $:
\begin{gather}
	 \phi^{lines}_{m}(t) = \phi_{CW}(t) + m\phi_{RF}(t). \label{eq:SM_normal_EO}
\end{gather}
That is, the phase noise of a standard EO comb follows the standard model by having the same form, with $ \phi_0(t)$ = $\phi_{CW}(t) $ being the phase noise that is common to all lines and $ \phi_{rep}(t)=\phi_{RF}(t) $ being the repetition rate phase noise.

\subsection{Resonant EO combs} \label{sec:theory_resonant_eo}
The resonant EO comb complicates this picture with the addition of a cavity around the phase modulator (see Fig. \ref{fig:setup_and_simulation}). The cavity allows the light to recirculate multiple times through the same modulator and causes it to interfere with itself from earlier round-trips. The cavity lifetime quantifies how long the light will survive inside the cavity before being coupled out. This section will discuss the role that this cavity lifetime plays in the interference that happens as the light recirculates inside the cavity. It will also discuss whether the cavity could cause deviations from the standard model of EO combs and what such deviations might look like. \vspace{0.2cm} 

Earlier work has shown that when the fluctuations of the external phase noise sources ($ \phi_{CW}(t) $ and $ \phi_{RF}(t) $) are slow, compared to the cavity lifetime, they transfer directly through the cavity and affect the comb lines similarly to the standard model of EO combs\cite{buscaino2020design,kim2017cavity}. We can motivate this direct transfer of slow phase fluctuations onto the comb lines by noting that light within the cavity escapes much faster than these fluctuations evolve. That is, if the characteristic timescale of the noise fluctuations is much longer than the cavity lifetime, the cavity will be close to transparent for these slow fluctuations. As an example, consider fluctuations in the CW laser's phase noise with a frequency around 1 kHz going into a cavity with a lifetime of $ 1 $ $ \si{\mu s} $. In this case, most of the light will have exited the cavity before these $ 1 $ kHz fluctuations have changed the incoming light's phase significantly. This provides an intuitive understanding of why the resonant EO comb, like a standard EO comb, is well-described by the standard model at long timescales/low frequencies.

However, for fluctuations that are fast compared to the cavity lifetime, a significant phase change will occur within one cavity lifetime. The light inside the cavity will carry these fluctuations long enough for interference to take place. This interference affects how the external phase noise sources are transferred to the comb lines and potentially causes a breakdown of the standard model. We will define the cavity linewidth $\Delta\nu_{cav}$ (inversely related to the cavity lifetime) as the characteristic frequency scale of this potential breakdown of the standard model:
\begin{gather}
	\Delta\nu_{cav} = f_{FSR}/F,
\end{gather}
where $ f_{FSR} $ is the free spectral range (FSR) of the cavity in Hz, and F is the (unit-less) finesse of the cavity. To represent the fast dynamics beyond this $ \Delta\nu_{cav} $, we introduce the function $ \mathcal{F}_{cav} $ such that the phase noise of the $m^{\rm th}$ comb line becomes:
\begin{gather}
	\phi^{lines}_{m}(t) = \phi_{CW}(t) + m\phi_{RF}(t) + \mathcal{F}_{cav}\big(m,\phi_{CW}(t),\dots\big). \label{eq:phi_lines_functional}
\end{gather}
$ \mathcal{F}_{cav} $ represents the phase noise dynamics of the cavity that go beyond the standard model. Its exact form is quite complicated, but it can be expressed as a function by writing each step in the simulation and signal processing as an analytical function and subtracting the standard model's contributions from $ \phi_{CW}(t) $ and $ \phi_{RF}(t) $. Due to the complexity of the function's expression, it is not informative to study it in detail. In this work, we instead test for such deviations to the standard model by simulating the effects of $ \mathcal{F}_{cav} $ and decomposing the simulated phase noise using subspace tracking\cite{razumov2023subspace}. Nevertheless, here are two observations on the properties of this function $ \mathcal{F}_{cav} $ that must be reflected in any noise components found beyond the standard model:
\begin{itemize}
	\item Following the discussion above, $ \mathcal{F}_{cav} $ must have a negligible contribution to the phase noise of the comb lines at long timescales/slow frequencies compared to the cavity's linewidth $ \Delta\nu_{cav} $.
	\item The contribution from $ \mathcal{F}_{cav} $ must depend only on the parameters of the cavity (finesse, insertion loss, propagation loss, and FSR), the modulator (modulation index $ \beta $, RF frequency), the CW laser (optical frequency, power), the comb line number, as well as the phase noise of the CW laser and RF generator. Other imperfections, such as vibrations in the cavity mirrors and other environmental effects are not included in this analysis to ensure the generality of our findings regarding the basic noise properties of resonant EO combs.
\end{itemize}

The simulations and experiments below will validate the need for additional noise components beyond the standard model to describe the effect from $ \mathcal{F}_{cav} $ on the comb lines' phase noise. Previous work investigating the resonant EO comb assumed slow phase noise fluctuations compared to $ \Delta\nu_{cav} $\cite{buscaino2020design}. Therefore they did not evaluate the impact of this term. Instead, they mainly discussed how the linewidth of each comb line changes due to the added complexity of a cavity around the phase modulator. To this end, they introduced the linewidth correction factor $ \chi_m(f) $, which describes how the optical spectrum of the CW laser changes as the modulation inside the cavity creates sidebands that constitute comb lines. Assuming the CW laser's linewidth is much ($ {\sim}10^4 $ times) lower than the cavity's FSR, it was observed that the correction to the linewidth is negligible. 

We extend their analysis by proposing to use their linewidth correction factor $ \chi_m(f) $ as an analytical description of how the comb line's phase noise PSDs are affected by the cavity. While this correction factor was originally defined on the optical spectrum of the CW light itself, here we instead propose to apply it to the CW laser's phase noise PSD ($ S_{CW}(f) $). This ad hoc change is mostly justified because we observe that it describes the effects of the cavity well. However, the change was originally motivated by the duality between two definitions of phase noise PSDs: one definition based on the phase fluctuations themselves at a certain frequency $ f $ in units of dB rad$ ^2/ $Hz, the other definition based on the comb's optical spectrum at an offset frequency $ f $ from the carrier with units of dBc$/$Hz \cite{rubiola2008phase}. While this duality is not always exact, it serves as a link between the PSD of the comb's optical spectrum (on which $ \chi_m(f) $ is defined) and the PSD describing the phase/frequency noise (which is the concern of our work). If no RF generator phase noise is present, the phase noise PSD of the $m^{\rm th}$ comb line as a function of Fourier frequency $ f $ is proposed to be:
\begin{gather}
	S_m(f) = \frac{|\chi_m(f)|^{2}}{|\chi_m(0)|^{2}} S_{CW}(f{-}mf_{RF}) \label{eq:lin_corr_applied},
\end{gather}
where it was used that the correction should vanish (i.e. become unity) at $ f=0  $ to normalize it. The unnormalized linewidth correction factor $ \chi_m $ is defined as (adapted from \cite{buscaino2020design}):
\begin{gather}
	\chi_m(f) = \sum_{n=1}^{\infty} r^n J_m(\beta n)e^{-j2\pi n f/f_{FSR}} \label{eq:chi_def},
\end{gather}
where $ r $ is the round trip field gain in the cavity,$ J_m $ is the $ m^{\rm th} $ order Bessel function, $ \beta $ is the modulation index, and $ f_{FSR} $ is the FSR of the cavity. Eq. \eqref{eq:lin_corr_applied} will be used as an analytical expression of how the phase/frequency noise PSDs of the comb lines are affected by the cavity and modulator of the resonant EO comb.

\subsection{Subspace tracking} \label{sec:subspace}
Subspace tracking seeks to find the underlying components that are responsible for the phase noise of all comb lines by looking at their correlation \cite{razumov2023subspace}. The central idea of this method is to write the phase noise of the comb lines in Eq. \eqref{eq:phi_lines_functional} as a linear combination of some $ P $ number of noise components. This is done using the $ M {\times }P $-dimensional generation matrix $ \mathbf{H} $:
\begin{gather}
	\boldsymbol{\phi}^{lines}(t) = \mathbf{H} \boldsymbol{\phi}^{comps}(t), \label{eq:generation_continuous}
\end{gather}
where $ \boldsymbol{\phi}^{lines}(t) $ is a vector containing the phase noise of each of the $ M $ comb lines of interest while $ \boldsymbol{\phi}^{comps}(t) $ contains the $ P $ phase noise components needed to describe these comb lines. In the case of an EO comb (resonant or not), where the first two components of $ \boldsymbol{\phi}^{comps} $ are the known contributions from the CW laser and RF generator, this matrix equation takes the form:

\begin{gather}\label{eq:comb_line_generation_written_ouit}
	\begin{bmatrix}
		\phi^{lines}_{-\frac{M-1}{2}}(t) \\
		\vdots\\
		\phi^{lines}_{-1}(t) \\
		\phi^{lines}_{0}(t) \\
		\phi^{lines}_{+1}(t) \\
		\vdots \\
		\phi^{lines}_{+ \frac{M-1}{2}}(t)
	\end{bmatrix}
	=
	\begin{bmatrix}\,
		\begin{pmatrix}
			1 \\
			\vdots\\
			1	\\
			1	\\
			1	\\
			\vdots\\
			1
		\end{pmatrix},&\hspace{-5pt}
		\begin{pmatrix}
			{-}\frac{M-1}{2}\\
			\vdots\\
			{-}1    \\
			0     \\
			{+}1     \\
			\vdots\\
			{+}\frac{M-1}{2}
		\end{pmatrix},
		\mathbf{h_3},&
		\hspace{-4pt}\dots\,,&
		\hspace{-4pt}\mathbf{h_P}\,
	\end{bmatrix}
	\begin{bmatrix}
		\phi_{CW}(t) \\
		\phi_{RF}(t) \\
		\phi^{comps}_{3}(t) \\
		\vdots \\
		\phi^{comps}_{P}(t)
	\end{bmatrix},
\end{gather}
where $ \phi^{lines}_m $ is the phase noise of the $m^{\rm th}$ comb line. $ \mathbf{h}_p $ is the $ p^{\rm th} $ column-vector in the generation matrix with $ \mathbf{h}_1 $ consisting solely of $ 1 $'s reflecting the CW laser's contribution that is common to all lines and $ \mathbf{h}_2 $'s elements increasing in increments of $ 1 $ away from the central line ($ m{=}0 $) reflecting the linearly increasing contribution from the RF generator's phase noise. $ \phi^{comps}_p $ is the corresponding $ p^{\rm th} $ component, with $ p{=}1 $ and $ p{=}2 $ labelled explicitly as the phase noise contributions from the CW laser and RF generator. The higher order components ($ p{>}2 $) are not needed in a standard EO comb, but their inclusion will allow us to describe deviations from the standard model such as measurement noise or additional noise components introduced by the fast dynamics of the cavity represented by $ \mathcal{F}_{cav} $ in Eq. \eqref{eq:phi_lines_functional}. Throughout this work, we will refer back to these three concepts: the comb lines (that we directly measure), the generation matrix (where the $ p^{\rm th} $ column vector describes the $ p^{\rm th} $ component's effect on each comb line), and the components (whose contributions make up the comb lines' phase noise).

Plots involving the noise components and comb lines will show their frequency noise (FN) power spectral densities (PSD). This is chosen, as opposed to phase noise PSDs, so that any deviation from a purely Lorentzian spectral line shape is immediately apparent from the plots. A Lorentzian line shape will correspond to a flat FN PSD. Frequency noise is here defined as the temporal derivative of phase noise. To get the FN PSD this corresponds to multiplying the phase noise PSD by $ f^2 $. This makes the $ {\propto} f^{-2}$ Lorentzian phase noise flat $ {\propto} f^{0}$ frequency noise and will make measurement noise proportional to $ {\propto} f^{2}$ in frequency noise PSDs.

\section{Simulation} \label{sec:simulation}
In this section, we present a simulation of the resonant EO comb and its noise sources. By simulating each round trip in the cavity, we capture the fast dynamics and extract the phase noise of multiple comb lines simultaneously, revealing noise correlations that deviate from the standard model. The simulations allow us to study the impact of individual external noise sources as well as cavity properties such as mirror reflectivity and the cavity's free spectral range (FSR). The ultimate goal of this simulation is to gain insights into the phase noise of the comb lines, specifically by identifying noise components beyond those predicted by the standard phase noise model of frequency combs.

\begin{figure} [t]
	\centering
	\includegraphics[width=1\linewidth]{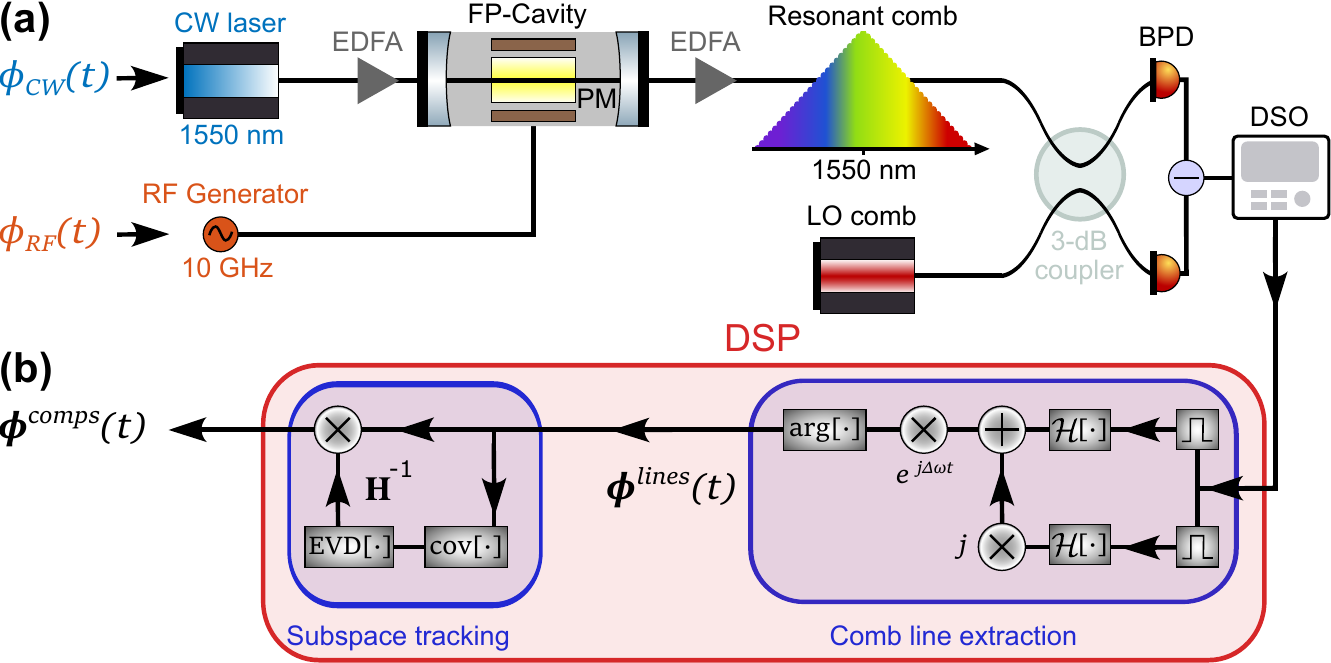}
	\caption{Simulated/experimental setup and digital signal processing (DSP). \textbf{(a):} $ \phi_{CW} $ and $ \phi_{RF} $ are the two known noise sources. They originate from the continuous-wave (CW) laser seeding the Fabry-Perot (FP) cavity and the radio-frequency (RF) generator driving the phase modulator (PM) placed inside the cavity. Erbium--doped fiber amplifiers (EDFA) are employed to amplify the signal (not included in simulation). The resulting comb is heterodyned with a local oscillator (LO) comb. The light enters a balanced photodetector (BPD) connected to a digital-storage-oscilloscope (DSO). \textbf{(b):} DSP algorithm. The phase noise of each comb line is extracted by employing the Hilbert transform (hil) to get both quadratures, applying a bandpass filter around the given comb line, adding the quadratures to reconstruct the complex electromagnetic field, detrending using the beat frequency $ \Delta\omega $, and lastly finding the phase using the argument (arg) function. The generation matrix is found by doing an eigenvalue decomposition (EVD) on the covariance (cov) matrix of the comb lines. The inverse generation matrix is applied on the comb lines to find the phase noise components.}
	\label{fig:setup_and_simulation}
\end{figure}
\subsection{Simulation methods}
The numerical simulation is performed according to Fig. \ref{fig:setup_and_simulation}. The continuous wave (CW) laser that seeds the resonant cavity is represented by a complex electric field with a wavelength of $1550\,$nm and phase noise $ \phi_{CW} $ with a Lorentzian linewidth of $100\,\text{kHz}$. Simulation of each round trip within the cavity involves injection of light, application of a phase modulation at $10\, \text{GHz}$, and extraction of a $ \sqrt{1{-}R} $ fraction of the electric field, where $R=0.94$ is the reflectivity of the mirrors. As the number of round trips increases, more and more comb lines are populated. To ensure the cavity has reached a steady state, the simulation is first run without extracting phase noise until only a $10^{-15}$ fraction of the initial round trip light remains. The RF signal for the modulator features Lorentzian phase noise generated with a linewidth of $ 100\,\text{Hz} $. While not a fully accurate representation of a real-world RF generator's phase noise, it is used here as a process that is easy to simulate and recognize. The modulation index, unless otherwise stated, is $ \beta = 0.05\,\pi $. The simulation is performed using a time resolution of $ 1\,$ps, ensuring that the cavity dynamics will be captured while avoiding aliasing effects.\vspace{0.2cm} 

The resulting E-field describing the comb is down-converted by combining it with a noiseless local oscillator (LO) comb in balanced photodetection. The phase noise of $ M=9 $ comb lines is extracted by applying a bandpass filter around each line. This enables simultaneous detection of multiple lines. Subspace tracking is then performed to investigate the comb line's phase noise correlations by performing the eigenvalue decomposition of their covariance matrix. This produces $ P $ significant eigenvectors, which make up the generation matrix $ \mathbf{H} $. The pseudo-inverse generation matrix is used to construct the corresponding phase noise components (for more information on the method see \cite{razumov2023subspace}).

\subsection{Simulation Results}
\begin{figure} [htpb]
	\centering
	\includegraphics[width=1\linewidth]{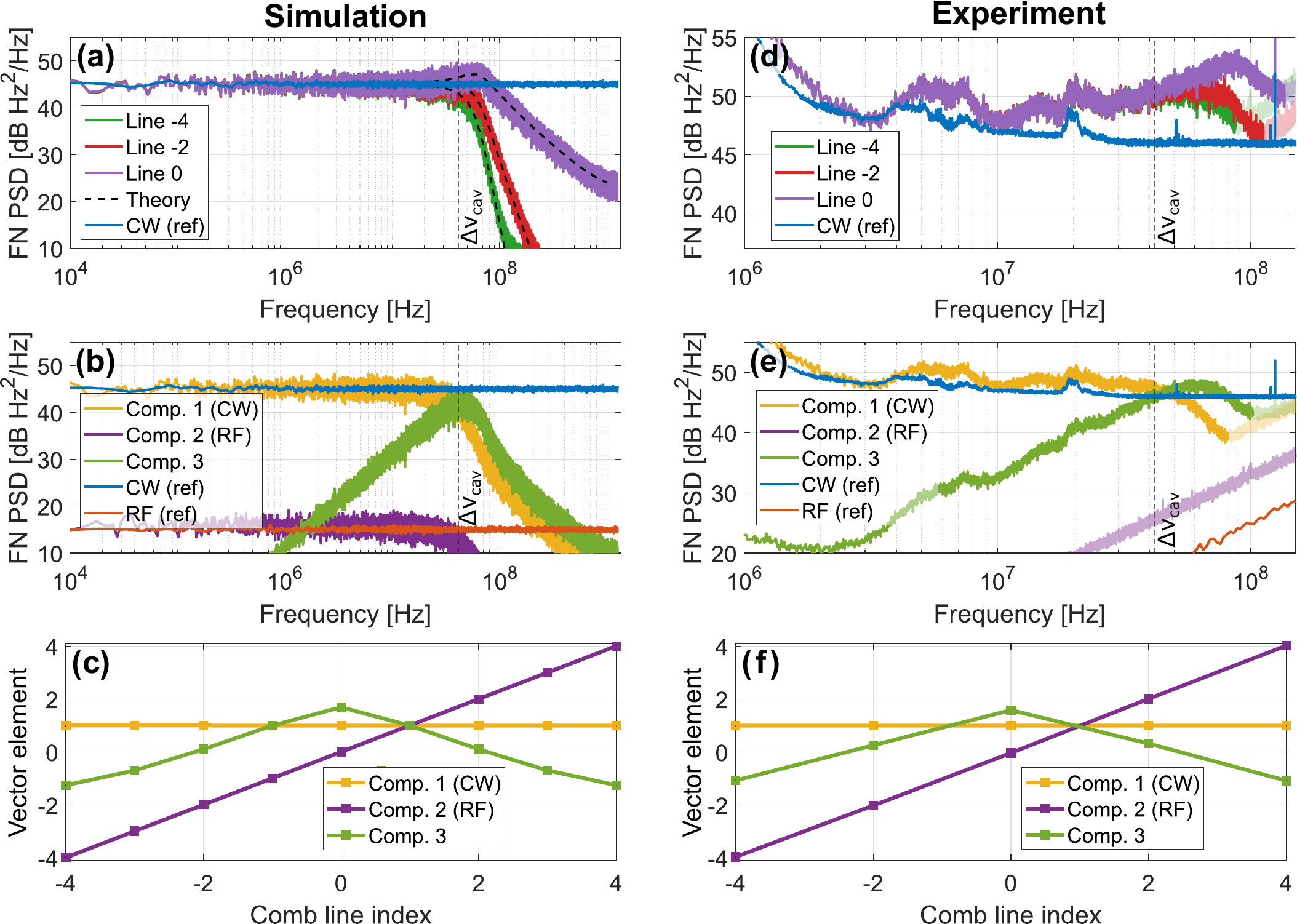}
	\caption{Simulation (left) and experiment (right) showing the high-frequency behaviour of frequency noise (FN) in resonant EO combs. Independent FN PSDs reference measurements are shown for the CW laser (blue) and RF generator (orange). \textbf{(a):} FN PSDs of the simulated comb lines. The theoretical prediction from Eq. \eqref{eq:lin_corr_applied} is shown in black dashed lines. \textbf{(b):} FN PSD of the noise components found by subspace tracking. \textbf{(c):} Matrix elements of vectors ($ \mathbf{h_P} $) in the generation matrix associated with found components. \textbf{(d):} FN PSDs of the experimental comb lines. As the PSD's become limited by measurement noise, they are marked by transparency. \textbf{(e):} FN PSDs of the experimental components found by subspace tracking. \textbf{(f)}: Matrix elements of vectors ($ \mathbf{h_P} $) in the generation matrix associated with found components.}
	\label{fig:exp}
\end{figure}
In this section, we will show and discuss the FN PSDs of the comb lines and components together with the vectors of the generation matrix that relate the two. As mentioned in Sec. \ref{sec:subspace}, the results are shown using FN PSDs such that the simulated Lorentzian phase noise appears flat. The FN PSDs of comb line $ 0 $, $ 2 $, and $ 4 $ are shown in Fig. \ref{fig:exp}(a). At low Fourier frequencies compared to $ \Delta\nu_{cav} $, the comb line PSDs all overlap and align with the true reference PSD of the CW laser. The RF generator's contribution, that increases linearly with comb line number, is not visible. This is because the simulated parameters where chosen to match the experiments, where the RF generator's phase noise is much lower than that of the CW laser. Beyond the cavity linewidth $ \Delta\nu_{cav} = f_{FSR}/F\approx 40$ MHz, the comb lines no longer follow the CW laser's FN PSD. Phase/frequency noise beyond $ \Delta\nu_{cav} $ is fast enough to experience significant interference within the cavity lifetime.
This interference causes the comb line PSDs to exhibit a small peak before they are filtered below the reference for the CW laser. The central lines exhibit a larger peak in the PSD while the outer lines experience more noise filtering, possibly because they are generated after more round trips inside the cavity. As the comb lines no longer follow the simple linear relation from Eq. \ref{eq:SM_normal_EO}, this constitutes a breakdown of the standard phase noise model of frequency combs. The noise components found using subspace tracking will reveal insights about this deviation from the standard model and how these noise components are distributed over comb lines.\vspace{0.2cm} 

The FN PSDs of these components are shown in Fig. \ref{fig:exp}(b). The first two components align with the true references for CW laser (blue) and RF generator (orange) phase noise up until the breakdown of the standard model at $ \Delta\nu_{cav} $. After this, the two components are filtered below their reference values, meaning that fast noise fluctuations entering the cavity are filtered away. This filtering is an expected effect coming from the transfer function of the cavity \cite{kim2017cavity,bienfang2001phase}. However, a 3rd component (green) is seen increasing until it becomes dominant at higher frequencies. The 3rd originates due to the fast dynamics of the cavity represented by $ \mathcal{F}_{cav} $ in Eq. \eqref{eq:phi_lines_functional} and, as predicted in Sec. \ref{sec:theory_resonant_eo}, this contribution is negligible for frequencies that are much slower than $ \Delta\nu_{cav} $. To preserve the legibility of the figures, higher-order components ($ p{>}3 $) have been excluded. Their contribution to the comb lines is smaller than the first three. Fig. \ref{fig:exp}c shows the column vectors $ \mathbf{h}_p $ of the generation matrix in Eq. \eqref{eq:comb_line_generation_written_ouit}. These column vectors corresponds to the noise components discussed above. Component 1 (yellow) has a flat contribution over comb lines, which agrees with $ \phi_{CW} $ in the standard model. Similarly, the linearly increasing component 2 (purple) aligns with the theoretical prediction for the RF generator's contribution $ \phi_{RF} $. The 3rd component (green) breaks the standard model by having a maximal contribution to the central lines that lowers for lines further out. \vspace{0.2cm} 

As an analytical description of these effects, Eq. \eqref{eq:lin_corr_applied} offers a description of how the FN PSD of the comb lines are affected by the cavity. The analytically predicted FN PSDs are shown as black dashed lines in Fig. \ref{fig:exp}(a) and agree excellently with the the extracted PSDs from the simulation. This agreement between the analytical expression and numerical results indicates that the deviation from the standard model of frequency combs can be accounted for by modelling the transfer of the CW laser's phase noise through a FP-cavity with modulation. 

\section{Experimental verification}
\subsection{Experimental methods}
The experiment is performed following the setup in Fig. \ref{fig:setup_and_simulation}(a). A $ 1550 $ nm DFB laser ($ 1 $ MHz reported full width at half maximum linewidth) is fed into the $ 2.5 $ GHz-FSR FP-cavity with a Finesse of 58\cite{xiao2008toward}. The phase-modulator is driven by a $ 10 $ GHz RF signal generated from a Synergy phase-locked oscillator (model: KDFLOD1000-8). Unless otherwise stated, the modulation index is $ \beta=0.05\,\pi $. Independent measurements of the CW laser and RF generator's frequency noise PSDs are shown in Fig. \ref{fig:exp}(e). The RF generator exhibits very low frequency noise at high frequencies, resulting in the reference measurement being dominated by measurement noise within the shown frequency range. 

The cavity length is locked by applying a DC voltage to a micro-strip inside the lithium niobate modulator in accordance with the Pound-Drever-Hall technique (see supplementary material of \cite{sekhar202320} for more details). \vspace{0.2cm} 

The output spectrum from the cavity is heterodyned with a standard $ 20 $ GHz EO comb with lower noise properties (CW laser linewidth is below $ 1 $ kHz and RF generator's FN PSD is less than that the resonant comb's RF generator at low frequencies) in a $ 5 $ GHz BDX balanced detector from Thorlabs. Thus every second line of the $ 10 $ GHz resonant EO comb is down-converted. The signal is digitized in a $ 2$ GHz bandwidth digital-storage oscilloscope. Signal processing follows the same procedure as the above numerical simulation and is shown in Fig. \ref{fig:setup_and_simulation}. 

\subsection{Experimental results}
Fig. \ref{fig:exp} compares the simulated (left) with the experimental (right) results. The extracted comb lines on Fig. \ref{fig:exp}(d) is in reasonable agreement with the simulated ones. Similar to the simulation, the comb lines should follow the reference measurement of the CW laser's FN PSD at low frequencies compared to the cavity linewidth $ \Delta\nu_{cav}$. The authors believe that the observed disagreement arises from the drifting noise properties of the DFB laser. At high frequencies, compared to the cavity linewidth, the lines feature a peak in PSD familiar from the simulated lines. Again, the central line experiences a larger noise peak, while the lines further away from the center are presumably more filtered. However, this filtering is difficult to observe, as the comb lines are limited by $ \propto f^2 $ measurement noise at the highest frequencies (indicated by transparent PSDs). This comes from the detector picking up amplified spontaneous emission originating in the EDFA's.

The decomposition into noise components from subspace tracking is similar between experiment and simulation. Fig. \ref{fig:exp}(e) shows that the 1st component (yellow) follows the CW laser's reference up until the $ \Delta\nu_{cav} $, the characteristic frequency scale for the breakdown of the standard model. Component 2 (purple), corresponding to the RF generator's contribution, is completely limited by measurement noise. This is because the RF generator has very low frequency noise at the shown timescales (the RF reference measurement is also limited by measurement noise here). At longer timescales/lower frequencies than shown in the Fig. \ref{fig:exp}(e), the RF generator has a significant contribution which makes it the 2nd-most significant component. The 3rd component (green) increases towards the higher frequencies before peaking at similar values to the simulation. Features in the 3rd component's PSD in around $ 1 $ MHz, $ 6 $ MHz and $ 20 $ MHz resemble those of the 1st component in Fig. \ref{fig:exp}(e), indicating that the 3rd component is related to the CW laser's noise. This would agree with the analytical description of the deviation from the standard model originating from the CW laser's phase noise interfering with itself inside the cavity with phase modulation. Looking at the corresponding vectors in Fig. \ref{fig:exp}(f), it is clear that the 3rd component is distributed differently over comb lines constituting a deviation from the standard phase noise model of frequency combs. Component 1 and 2 are the familiar constant and linearly increasing contributions from the standard model, but the 3rd component again peaks around comb line 0, with a decreasing matrix element for comb lines further out. \vspace{0.2cm} 

The location of this 3rd component's peak in the frequency domain is investigated by analyzing the PSDs of the comb lines from Eq. \eqref{eq:lin_corr_applied}. Assuming that the 3rd component is responsible for the observed peak in the FN PSDs of the comb lines, the 3rd component's peak $ f_3 $ can be found by identifying the peak in the analytical comb line PSDs. Numerically extracting this maxima reveals an approximate relationship (which becomes more accurate for cavities with higher finesse) between the $ f_3 $ and the modulation index $ \beta $ :
\begin{gather}
	\frac{f_3}{f_{FSR}} \approx \frac{1}{2}\frac{\beta}{\pi}. \label{eq:f3_beta}
\end{gather}
\begin{figure} [htpb]
	\centering
	\includegraphics[width=1\linewidth]{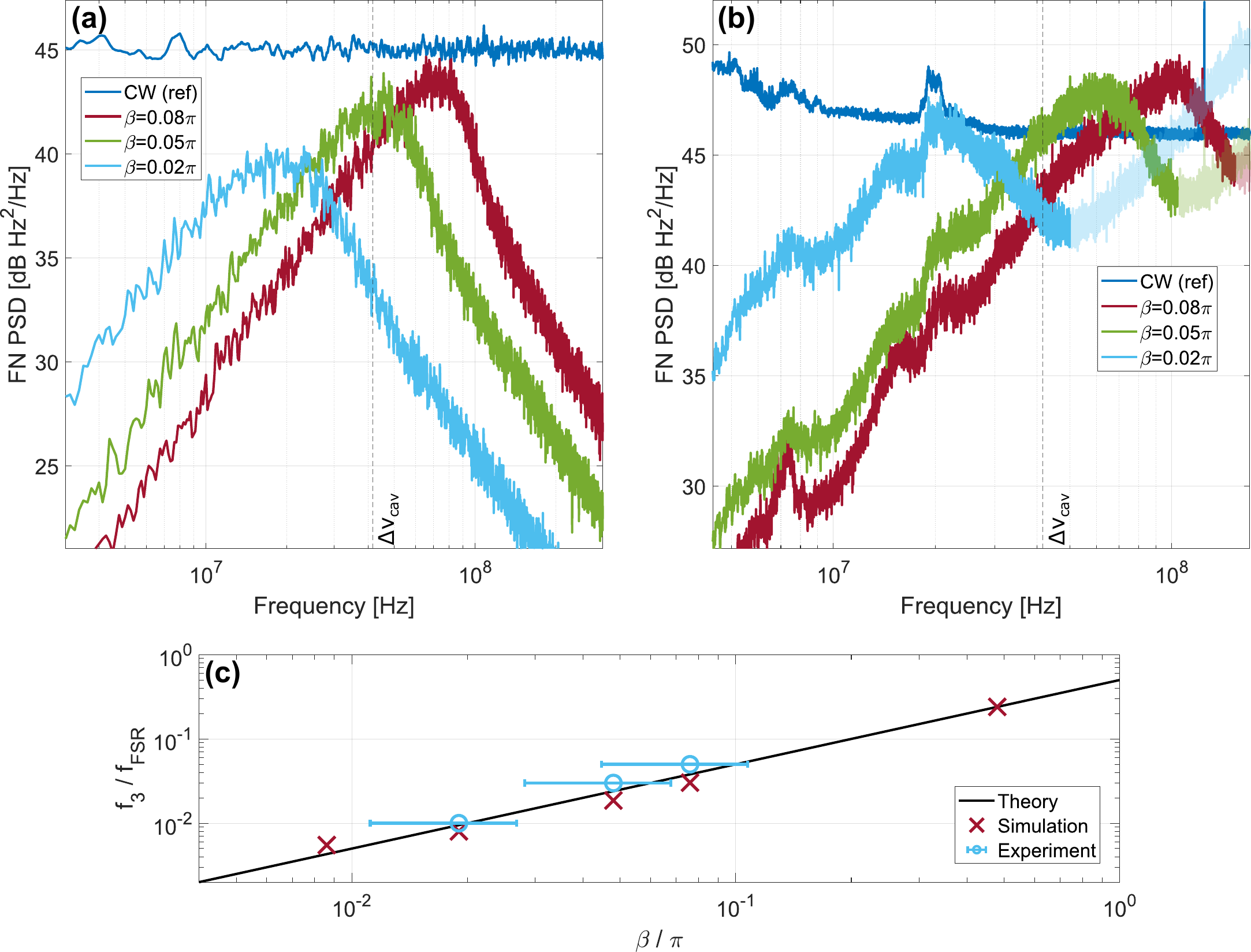}
	\caption{3rd component as RF power is swept. \textbf{(a):} Simulation \textbf{(b):} Experiment with matching parameters. \textbf{(c):} Peak of 3rd component for different values of beta in simulation and experiment compared to the theoretical prediction from Eq. \eqref{eq:f3_beta}.}
	\label{fig:rf_power}
\end{figure}
To verify this relation, a range of $ \beta $s is tested experimentally and numerically. The resulting PSDs of the 3rd component are shown in Fig. \ref{fig:rf_power}(a) and \ref{fig:rf_power}(b). There is good agreement between the simulation and experiment, with stronger modulation indices leading to a higher peak in the FN PSD and a shift to higher frequencies of the peak's location. Fig. \ref{fig:rf_power}(c) collects the analytical prediction with the numerical and experimental results for the peak's location. The error bars on the experimental results come from the uncertainty in the exact loss of the RF cables leading to the phase modulator. The approximate analytical relation in Eq. \eqref{eq:f3_beta} seems to match the numerical and experimental results.\vspace{0.2cm} 

The increase of the 3rd component with modulation index is problematic as it will reduce the cavity's filtering of the comb lines' phase noise. Previous applications of resonant EO combs feature $ \beta $s between $ 0.3\pi$ and $ 1.2\pi  $ \cite{kim2017cavity,sekhar202320,zhang2019broadband}, which are significantly larger values than those of the shown simulations and experiments. Consequently, the practical need for a second dedicated filter cavity could be explained by a reduction in noise filtering due to this 3rd component.

The $ \beta $s presented so far in this work were chosen to bring the peak of the 3rd component inside the experimentally limited effective Nyquist frequency. This effective Nyquist frequency is determined by the bandwidth of the bandpass filters used to extract the comb lines. With a larger oscilloscope bandwidth, the lines could be spaced out further allowing a wider bandpass filter when extracting each line. 

In real-world applications higher $ \beta $s are chosen to ensure a broad comb. $ \beta $ approximately affects the power of the $ m^{\rm th} $ comb line $ P_m $ exponentially\cite{kourogi1993wide}:
\begin{gather}
	P_m \propto e^{\frac{-|m|\pi}{\beta F}}, \label{eq:exp_decay_line_power}
\end{gather}
where $ F $ is the finesse of the cavity. This relation hints at a trade-off between the finesse of the cavity and the modulation index of the EO-modulator. Any reduction in the modulation index $ \beta $ to reduce the effect of the 3rd component will need to be met with a corresponding increase in the cavity finesse $ F $ to maintain a broad comb bandwidth. The following section will show simulations with higher $ \beta $ and discuss how this 3rd component affects resonant EO-combs designed for applications. 

\section{Implications for design}
As the 3rd component is a deviation from the standard model, it could be problematic for experimental phase-stabilization mechanisms and digital compensation algorithms that rely on the standard model to reduce the phase noise of frequency combs\cite{papp2014microresonator,ideguchi2014adaptive,burghoff2019generalized,lundberg2020phase}. It is, therefore, important to choose design parameters so as to minimize it. 

\begin{figure} [htpb]
	\centering
	\includegraphics[width=1\linewidth]{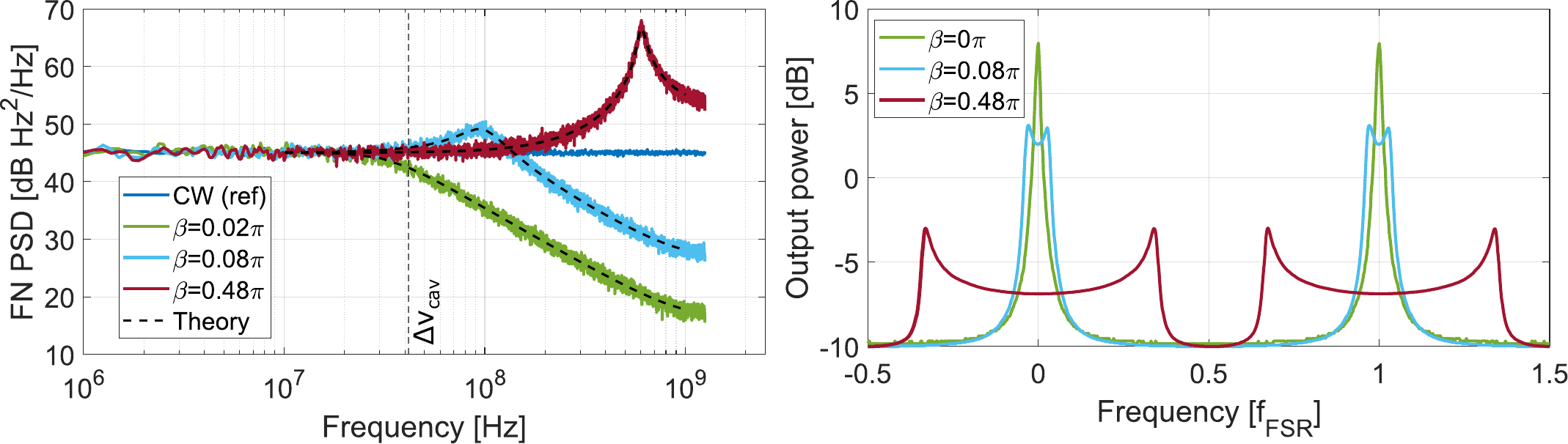}
	\caption{\textbf{Left:} Simulated frequency noise PSD of comb line 0. \textbf{Right:} Experimental transfer function for varying modulation indices. x-axis is normalized to the FSR of the cavity $ f_{FSR}=2.5$ GHz.}
	\label{fig:line0_and_transfer}
\end{figure}

Another problem created by the 3rd component is that the appealing noise filtering effect from the resonant cavity is limited by its presence. Earlier work has observed the need for a second external filter cavity whose FSR matches the comb's repetition rate before supercontinuum generation \cite{sekhar202320} to reduce broadband noise. This need implies a reduced filtering of broadband phase and amplitude noise at large modulation indices in resonant EO combs, reducing their advantage over standard EO combs. As a possible explanation, the central comb line is plotted for increasing modulation indices in Fig. \ref{fig:line0_and_transfer}(a). At a low modulation index of $ \beta=0.02\pi $ the phase noise is greatly reduced compared to the reference of the CW laser. While for a larger modulation index of $ \beta=0.48\pi $, the 3rd component brings the noise strictly above the CW laser's FN PSD. \vspace{0.2cm} 

From a physical point of view, these effects could be due to the modulation inside the cavity changing its effective transfer function. The modulator adds a varying phase shift to the light, making various frequencies resonant. The total transfer function was measured experimentally, following \cite{xiao2008toward} for different modulation indices and is shown in Fig. \ref{fig:line0_and_transfer}(b). The split resonance comes from the sinusoidal modulation which spends more time in the maximal and minimal phase shifts of $ \pm\beta $. The resemblance between the cavity transfer function and the theoretical prediction from Eq. \eqref{eq:lin_corr_applied} indicates that the peak induced by the 3rd component could be caused by the cavity transfer function in the presence of modulation. 

The modified transfer function in the presence of modulation filters fluctuations of the E-field such that they are increased relative to their value at the carrier. This increase will cause a growth in the phase noise of the comb lines that peaks around the maximal phase shift from the modulation, which is $ \beta/\pi $. This peak is consistent with the 3rd component peak discussed in this work. It is still unclear why that increase from the transfer function would have a different distribution over comb lines. It could arise from the different comb lines having a different lifetime inside the cavity, causing them to experience the filtering from the cavity for different lengths of time.\vspace{0.2cm} 

\begin{figure} [htpb]
	\centering
	\includegraphics[width=0.5\linewidth]{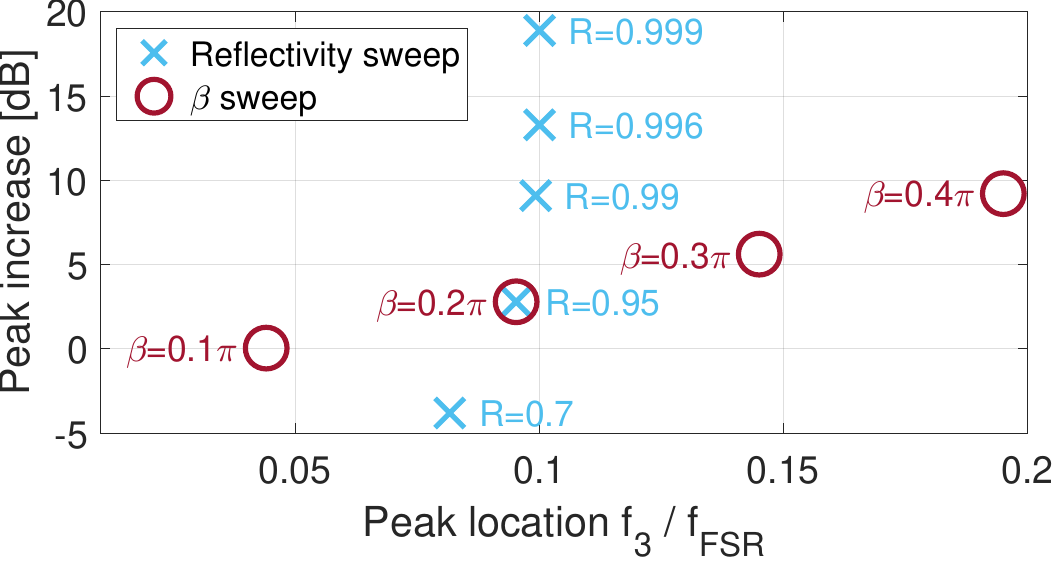}
	\caption{Simulation - Peak increase in FN PSD of 3rd component relative to CW laser's PSD vs peak location in units of $ f_{FSR} $. Two different sweeps are shown in the  reflectivity $ R $ and applied modulation index $ \beta $ respectively. Unless specified, the markers feature $ \beta=0.2\pi $ and $ R=0.95 $.}
	\label{fig:S3_vs_f3}
\end{figure}

By optimizing the design of resonant EO combs, these problematic effects of increased phase noise and reduced filtering of broadband noise can be avoided. Two important parameters for the resonant cavity are the modulation index $ \beta $ and the finesse F, governed by the reflectivity $ R $ of the cavity's mirrors. These two parameters determine the exponential decay of the comb lines' power in Eq. \ref{eq:exp_decay_line_power} and thus the bandwidth of the comb \cite{kourogi1993wide,buscaino2020design}. To maintain a fixed bandwidth one can choose to reduce one parameter while increasing the other. Thus tuning these parameters offers a way to limit the effect of the 3rd component, given target comb bandwidth \cite{heeboell2024resonant}. Fig. \ref{fig:S3_vs_f3} illustrates this trade-off by plotting the location of the 3rd component peak against the increase in its FN PSD relative to the flat CW laser's FN PSD. A range of simulated modulation indices and reflectivities are plotted. Increasing the reflectivity leads to a large increase in the peak FN PSD while increasing the modulation index primarily shifts the peak location to higher frequencies. This offers some insight into the design of resonant EO combs. For example, it might make sense to lower the reflectivity to reduce the peak of the 3rd component while increasing $ \beta $ to move it to higher frequencies beyond scope of the experiment. Conversely, lowering $ \beta $ could move the 3rd peak down to lower frequencies allowing broadband noise to be filtered by the built-in resonant cavity. 
 
\section{Conclusion}
Although resonant EO combs are found to follow the standard model of frequency combs' phase noise at low frequencies, an extension is needed for frequencies beyond the cavity linewidth. In addition to the first two components from the standard model ($ \phi_{CW} $ and $ \phi_{RF} $), a 3rd significant component is first estimated analytically and then confirmed numerically and experimentally using subspace tracking. This 3rd component induces a peak in the frequency noise power spectral density of the comb lines. A simple analytical relation between the peak's location and the modulation index is presented and confirmed both in simulation and experiment. To the best of the authors knowledge, this is the first well-described phase noise effect beyond the two known components from the standard model. Thus showing that subspace tracking can be used to identify residual noise which could be difficult to suppress. Beyond the analytical expression, a physically intuition is proposed based on the transfer function of the cavity in the presence of modulation. This offers an explanation for the missing noise filtering anticipated in resonant EO combs. Depending on the application of interest, one can optimize the impairment from this 3rd component by choosing the reflectivity of the cavity's mirrors and the modulation index applied inside the cavity. The methods employed in this work can similarly be applied to other types of frequency combs to determine whether they exhibit phase noise beyond the standard model, and to assist in optimizing their design parameters to limit these effects.


\medskip
\noindent\textbf{Funding.} This work was supported by the SPOC Centre (DNRF 123), the Villum Foundation (VYI OPTIC-AI grant no. VIL29344 and VI-POPCOM grant no. VIL54486) and DARPA MTO PIPES contract. The work at University of Colorado was supported by NSF AST 2009982 and NSF QLCI Award No. OMA-2016244. Matt Heyrich acknowledges support from the NSF Graduate Research Fellowship Program

\medskip
\noindent\textbf{Data availability.} Data underlying the results presented in this paper may be obtained from the authors upon reasonable request

\medskip
\noindent\textbf{Disclosures.} The authors declare no conflicts of interest.


\bibliography{references}

\end{document}